\documentclass[aip,reprint]{revtex4-1}
\usepackage{amsfonts,amsmath,amssymb}
\usepackage[squaren]{SIunits}
\usepackage[usenames,dvipsnames]{color}
\usepackage[pdftex]{graphicx}
\usepackage{bm,bbm}
\usepackage{hyperref}

\definecolor{mygreen}{rgb}{0,0.5,0} 
\definecolor{myblue}{rgb}{0,0,0.75} 
\definecolor{mymagenta}{cmyk}{0,1,0,0.12}

\newcommand{\var}{\mathrm{var}}
\newcommand{\cov}{\mathrm{cov}}

\newcommand{\width}{L}
\newcommand{\Bunit}{\hat{B}}
\newcommand{\bBperp}{{\bf B}_{\perp}'}
\newcommand{\bBpar}{{\bf B}_{\parallel}'}

\newcommand{\bunk}{{\bf x}}

\newcommand{\bfe}{{\bf e}}

\newcommand{\x}{{\bf x}}
\newcommand{\y}{{\bf y}}

\draft 

\begin{document}

\title{Real-time vector field tracking with a cold-atom magnetometer}

\newcommand{\ICFOAddress}{ICFO -- Institut de Ciencies Fotoniques, Av. Carl Friedrich Gauss, 3, 08860 Castelldefels, Barcelona, Spain}
\newcommand{\ICREAAddress}{ICREA -- Instituci\'{o} Catalana de Re{c}erca i Estudis Avan\c{c}ats, 08015 Barcelona, Spain}
\newcommand{\CavendishAddress}{Cavendish Laboratory, University of Cambridge, JJ Thomson Avenue, Cambridge CB3 0HE, UK}
\newcommand{\OptosAddress}{Optos, Carnegie Campus, Dunfermline, KY11 8GR, Scotland, UK}

\author{N. Behbood}\affiliation{\ICFOAddress}
\author{ F. Martin Ciurana}\affiliation{\ICFOAddress}
\author{G. Colangelo}\affiliation{\ICFOAddress}
\author{M. Napolitano}\affiliation{\ICFOAddress}
\author{M.W.~Mitchell}\affiliation{\ICFOAddress}\affiliation{\ICREAAddress}
\author{R. J.~Sewell}\affiliation{\ICFOAddress}

\date{\today}

\begin{abstract}
We demonstrate a fast three-axis optical magnetometer using cold, optically-trapped $^{87}$Rb gas as a sensor.  By near-resonant Faraday rotation we record the free-induction decay following optical pumping to obtain the three field components and one gradient component.  A single measurement achieves shot-noise limited sub-nT sensitivity in \unit{1}{\milli\second}, with transverse spatial resolution of $\sim$\unit{20}{\micro\meter}.  We make a detailed analysis of the shot-noise-limited sensitivity.
\end{abstract}


\maketitle

\newcommand{\rootHz}{$\sqrt{{\rm Hz}}$}

Control of magnetic fields is critical to many applications, for example high-sensitivity instruments\cite{LudlowS2008} and atomic physics experiments\cite{DaiPRL2012}.  Hot-atom optical magnetometers\cite{BudkerNP2007} offer sub-fT sensitivity at the \unit{\centi\meter} scale\cite{BudkerNP2007} and sub-pT at the \unit{\milli\meter} scale\cite{SchwindtAPL2007}.  Cold atom magnetometers have demonstrated \unit{10}{\pico\tesla}  gradient sensitivity at the \unit{10}{\micro\meter} scale\cite{VengalattorePRL2007,KoschorreckAPL2011} and \unit{100}{\pico\tesla} gradient sensitivity at the  \unit{3}{\micro\meter} scale.\cite{WildermuthAPL2006}  A single-ion narrow-band ($2 \times 10^{-3}$~Hz) magnetometer\cite{KotlerN2011} showed \unit{10}{\pico\tesla} sensitivities.  Many of these systems are designed for sensing of one field or gradient component.  In contrast, precise control of fields requires simultaneous knowledge of all field components and possibly gradients as well.  Recent work with hot-atom magnetometers has demonstrated three-axis sensing and control with \unit{\nano\tesla} precision at~\unit{\centi\meter} scales,\cite{CoxPRA2011,FangRSI2012} while a three-axis modulated cold-atom magnetometer has shown nT sensitivity at \unit{300}{\micro\meter} scales.\cite{HaycockPRA1998,SmithJPB2011} 

Here we demonstrate sub-nT sensitivity at \unit{50}{\micro\meter} length-scale in a cold-atom magnetometer employing near-resonant Faraday rotation probing\cite{KubasikPRA2009}.  The instrument gives three-axis field information plus one gradient component, obtained by free-induction decay (FID) after optical pumping.
A full measurement requires \unit{1}{\milli\second}, and can be repeated with zero dead-time, allowing high-bandwidth recording and control of the vector field. No external magnetic fields need be applied, so the technique can be used for real-time monitoring during field-sensitive processes.  Our implementation is shot-noise limited and we give explicit expressions for its shot-noise-limited sensitivity. 

\begin{figure}[htbp]
\centering
\includegraphics[width=0.5\textwidth]{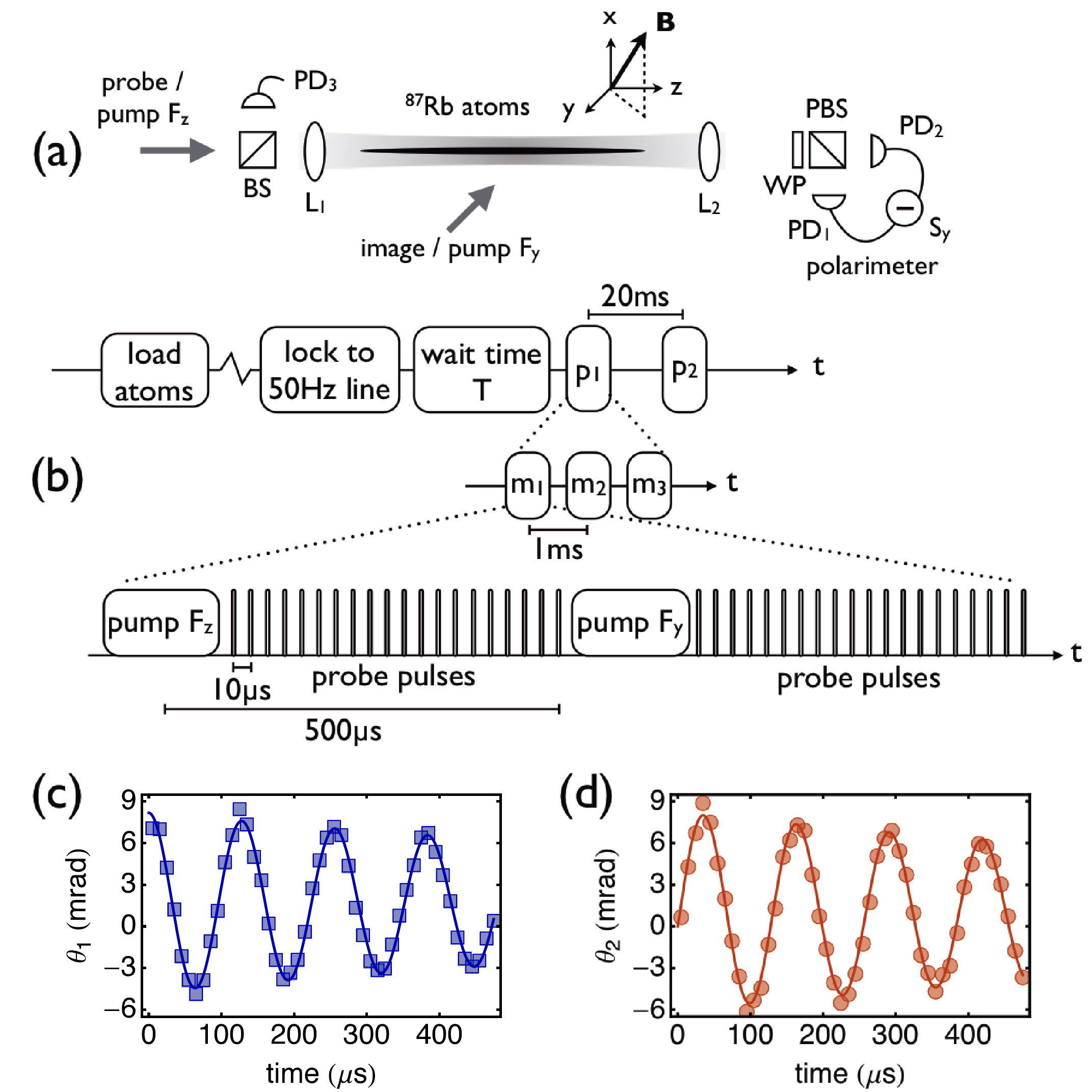}
\caption{
(Color online)
(a) Experimental geometry. 
PD: photodiode; L: lens; BS: beam-splitter; PBS: polarizing beam-splitter; WP: waveplate. 
(b) Schematic of the experimental sequence:
the atoms are initially polarized along $F_z$ via optical pumping and then probed with a sequence of \unit{1}{\micro\second} long pulses of light at \unit{10}{\micro\second} intervals.
The resulting FID signal is recorded over \unit{500}{\micro\second}.
We then immediately re--polarize the atoms along $F_y$ with an orthogonal optical pumping beam and record a second FID signal.
A single shot is thus acquired in \unit{1}{\milli\second}.
We fit the two measurements together to a simple model to extract the magnetic field components $B_i$.
Typical data from a single composite FID measurement are shown in (c) for an initially $F_z$-polarized state and (d) for an initially $F_y$-polarized state.}
For these data the field components extracted from the fit were ${\bf B}=\unit{(941, 310, 511)}{\nano\tesla}$ and the coherence time $\tau_c=\unit{1.3}{\milli\second}$.
\label{fig:Fig1}
\end{figure}

\newcommand{\ttwo}{T_2}
The experiment is shown schematically in Figure \ref{fig:Fig1}(a).  
An ensemble of atoms is held in an elongated optical trap, and subject to an unknown field ${\bf B}$.
The atoms are first optically pumped along the $z$ direction, so that the collective atomic spin ${\bf F}$ achieves the value $(0,0,F_z(0))$, and then allowed to precess around ${\bf B}$.  Off-resonance Faraday-rotation probing measures the rotation angle $\theta = G F_z$, where $G$ is a coupling constant, known from independent measurements.\cite{KubasikPRA2009}  We observe the FID signal\cite{Abragam}
\begin{equation}
\theta_1(t) = \frac{G}{|B|^2} \left[{B_z^2}+[{B_x^2+B_y^2}]\cos(\gamma |B| t) e^{-{t}/{\ttwo}} \right] F_z(0),
\label{eq:FIDSignalZStart}
\end{equation}
where $\gamma = \mu_B g_F / \hbar$ is the gyromagnetic ratio, $\mu_B$ is the Bohr magneton, $g_F$ is the Land\'{e} factor, and $\hbar$ is Planck's constant.  The transverse relaxation time $\ttwo = 1/(\width\gamma |B'_\parallel |)$ is due to the field-parallel gradient component $B'_\parallel \equiv \partial |{B}| /\partial_z$, and a Lorentzian distribution (full-width at half-maximum $\width$) of atoms along $z$, the trap axis.\cite{TaquinRPA1979}
The process is then repeated with the spins initially polarized $(0,F_y(0),0)$, to give
\begin{eqnarray}
\theta_2(t) &=& \frac{G}{|B|^2} 
	\left[ B_y B_z \left(1-\cos(\gamma |B| t)  e^{-{t}/{\ttwo}}\right) \right. \nonumber \\
	&&\quad + \left. B_x |B|\sin(\gamma B t)  e^{-{t}/{\ttwo}} \right] F_y(0).
\label{eq:FIDSignalYStart}
\end{eqnarray}

Fitting the composite signal from the two FID measurements gives  the three ${\bf B}$ components up to a global sign and $\ttwo$.  
The ambiguity can be lifted by applying a known field, if necessary.  
Representative traces are shown in Figure \ref{fig:Fig1}(c). 
Relative to other vector magnetometry techniques,\cite{CoxPRA2011,FangRSI2012,SmithJPB2011} this method is simple both in procedure and in interpretation and requires no applied B-fields, making it attractive for work with field-sensitive systems.\cite{DiazAguiloCQG2010,BlochNP2012}

To derive Equations (\ref{eq:FIDSignalZStart}) and  (\ref{eq:FIDSignalYStart}) we note that the microscopic spin operators evolve as ${\bf f}^i(t) = R(z_i, t) {\bf f}^i(0) $, 
where ${\bf f}^i$ is the spin of the $i$'th atom with position $z_i$ and $R(z, t) = \exp[\gamma_F  t |{\bf B}(z)| A_B]$, 
where
\begin{eqnarray}
A_B &\equiv&  \left(\begin{array}{ccc} 
0 & -\Bunit_z &\Bunit_y \\ 
 \Bunit_z & 0 & -\Bunit_x \\ 
-\Bunit_y & \Bunit_x & 0  \end{array} \right),
\end{eqnarray} 
is the generator of rotations about ${\bf B}$ and $\hat{\bf B} \equiv {\bf B}/|{\bf B}|$. 

Possible decoherence mechanisms include atomic motions and collisions, tensorial light shifts due to the probe light, and decoherence due to gradient of the field. 
In our experiment the effect of tensorial term of probe is negligible, since we are far detuned from $D_2$ transition line and we are using few photons for detection.
For the time-scales involved in this experiment,  decoherence due to collisions is negligible, whereas dephasing, i.e. differential precession due to field inhomogeneity, typically is not.   
In the language of magnetic resonance, we expect longitudinal relaxation to be much slower than transverse relaxation due to field inhomogeneity. 

Expanding the field as  ${\bf B}(z) \approx {\bf B}_0 + (\bBpar + \bBperp )  z + O(z^2)$, where $\bBpar$ is parallel to ${\bf B}_0$ and $\bBperp$ is perpendicular. 
We note that a change in the magnitude of ${\bf B}$ has an accumulating effect on the spin precession, i.e., the change in ${\bf f}$ grows with $t$.  
In contrast, a change in the  {direction} of ${\bf B}$ has a fixed effect:  
From the perspective of the measurement, a rotation of ${\bf B}$ is equivalent to a rotation of both the initial state and the measured component $F_z$.  
For small gradients $\partial_z {\bf B} \ll {\bf B}/l_{\rm atoms}$, where $l_{\rm atoms}$ is the length of the cloud, we can ignore $\bBperp$.  
This approximation, along with the fact that $A_B^{n+2} = - A_B^n$, allows us to write 
\begin{eqnarray}
R(z, t) 
 & \approx & \mathbb{I} + A_{B_0} \sin \omega(z) t + A_{B_0}^2 [1-\cos \omega(z) t ]
\end{eqnarray}
where $\omega(z) = \gamma_F |{\bf B}_0 + z \bBpar|$.

In our trap, we observe an atomic density $\rho(z)$ well approximated by a Lorenzian $\rho(z) =  \width / \pi(z^2 + \width^2)$ where $\width \approx 48$ {\micro m} is the full-width half-maximum extent of the ensemble. The collective spin ${\bf F} \equiv \sum_i f^i $ then evolves as
\begin{eqnarray}
{\bf F}(t) &=& \int dz \, \rho(z) R(z,t) {\bf f}(0) \\
& = & [\mathbb{I}\, + A_{B_0}^2 ] {\bf F}(0) \\
& & + e^{-\width \gamma_F |\bBpar| t} (A_{B_0}\sin  \omega_0 t - A_{B_0}^2\cos \omega_0 t  ) {\bf F}(0) \nonumber 
\end{eqnarray}
In the first term $\mathbb{I}\, + A_{B_0}^2$ describes a projector onto the direction of ${\bf B}_0$.  This is the steady-state polarization. The second line describes a decaying oscillation of the transverse components, i.e., those perpendicular to $\bf{B}_0$.

Our experimental apparatus has been described in detail elsewhere.\cite{KubasikPRA2009}
Briefly, we work with ensemble of up to $2 \times 10^5$ laser cooled $^{87}$Rb atoms in the $F=1$ hyperfine ground state. 
The atoms are held in a single-beam optical dipole trap with beam waist \unit{56}{\micro\meter}, which sets the minimum distance at which the field can be measured.  The atom cloud itself has \unit{20}{\micro\meter} lateral dimension, defining the transverse resolution.
The atoms are probed with \unit{\micro\second} duration pulses of linearly polarized light at \unit{10}{\micro\second} intervals, red detuned by $\unit{1.5}{\giga\hertz}$ from resonance with the $F=1\rightarrow F'=0$ transition of the D$_2$ line.
Each pulse contains on average \unit{10^7}{photons}.
After passing through the atoms, the light pulses are detected by a shot--noise--limited balanced polarimeter.\cite{WindpassingerMST2009} The experimental geometry is illustrated in Fig.~\ref{fig:Fig1}(a).

The detuning and photon number are chosen so that both probe--induced decoherence due to spontaneous emission and the perturbation due to tensorial light shifts are negligible during the measurement cycle.\cite{KubasikPRA2009} 
This allows us to use the simple model described by Equations~(\ref{eq:FIDSignalZStart}) and~(\ref{eq:FIDSignalYStart}) to fit the data.
We note that the measurement sensitivity could be increased by using more photons and/or reducing the detuning, at the cost of more elaborate data analysis.\cite{KubasikPRA2009,KoschorreckPRL2010b}


\begin{figure}[htbp]
\centering
\includegraphics[width=0.5\textwidth]{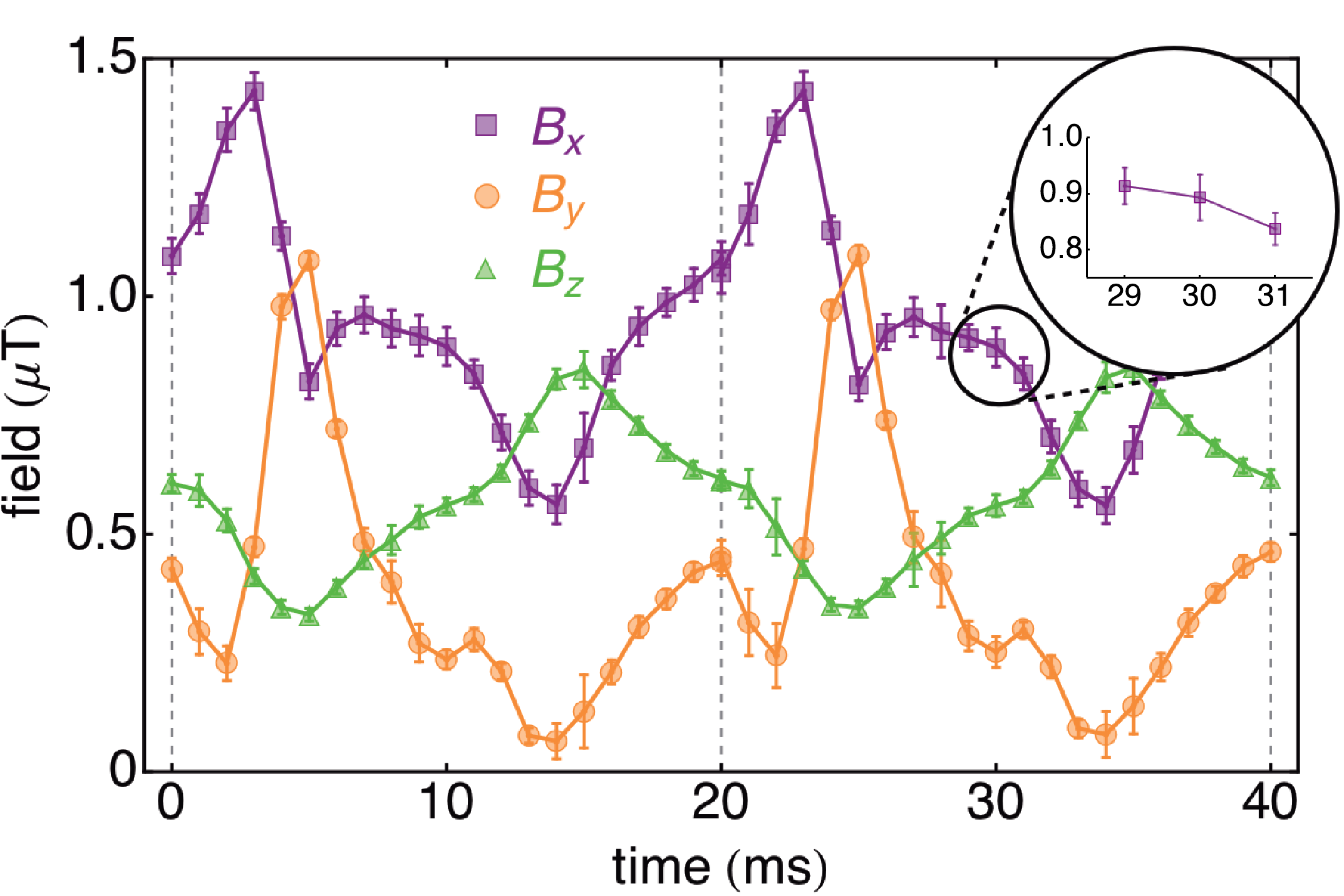}
\caption{
(Color online)
Recording of low frequency laboratory magnetic field noise, acquired with one sample point per cycle and varying phase with respect to the \unit{50}{\hertz} signal of the laboratory mains line, as in equivalent--time sampling with a digital oscilloscope.
We plot the field components estimated from the three measurement $m_j$ at each point $p_i$.
Inset: Three consecutive measurements are made at each point.
Here we plot the $B_x$ component estimated from the three measurements.
Error bars are $\pm1\sigma$ statistical errors.
\label{fig:equivalentTimeSample}}
\end{figure}

The initial atomic spin state is prepared via optical pumping with a single \unit{5}{\micro\second} duration circularly polarized pulse on resonance with the $F=1\rightarrow F'=1$ transition of the D$_2$ line and propagating either along the trap axis, i.e. the $z$-axis, to prepare an $F_z$-polarized state, or along the $y$-axis, to prepare an $F_y$-polarized state.
During the optical pumping, the atoms are uniformly illuminated with randomly polarized light on resonance with the $F=2\rightarrow F'=2$ transition of the D$_2$ line to prevent atoms accumulating the $F=2$ hyperfine state.
A single {\em composite FID measurement} consists of first preparing an $F_z$-polarized state and measuring the FID signal over \unit{500}{\micro\second}, then immediately preparing an $F_y$-polarized state and again making a FID measurement.  
A single shot is thus acquired in \unit{1}{\milli\second}.

To illustrate the technique, we first record the laboratory magnetic noise at the trap, shown in Figure~\ref{fig:equivalentTimeSample}.  Laser cooled atoms are first loaded into the dipole trap during \unit{\sim2}{\second} via a two--stage magneto--optical trap (MOT).  A small field is applied with three pairs of Helmholtz coils, and 
the experiment is  triggered on the \unit{50}{\hertz} signal of the laboratory mains line.  The field is sampled at a sequence of points $p_i$ at \unit{20}{\milli\second} intervals, with a variable wait time $T$ before the first point.  At each $p_i$ we make three consecutive composite FID measurements $m_j$, as shown in the inset of Figure~\ref{fig:equivalentTimeSample}.  The entire sequence is repeated 300 times to collect statistics. 

The results show good predictability of the field from one cycle to the next, with a typical statistical uncertainty of $\sigma_{B_i}=\unit{40}{\nano\tesla}$ for each field component.
Note that the experiment has no magnetic shielding, so that the observed variance is dominated by magnetic field noise from the laboratory environment.  We observe $\ttwo \approx \unit{1.5}{ms}$, which sets a limit on the coherence time and is important for design of future experiments.  The FID signals give information about one gradient component, the one along the average field.  With three FIDs, with applied bias fields along different directions, we can obtain $\partial {\bf B}/\partial z$, all the gradient components affecting the experiment.

\begin{figure}[htbp]
\centering
\includegraphics[width=0.5\textwidth]{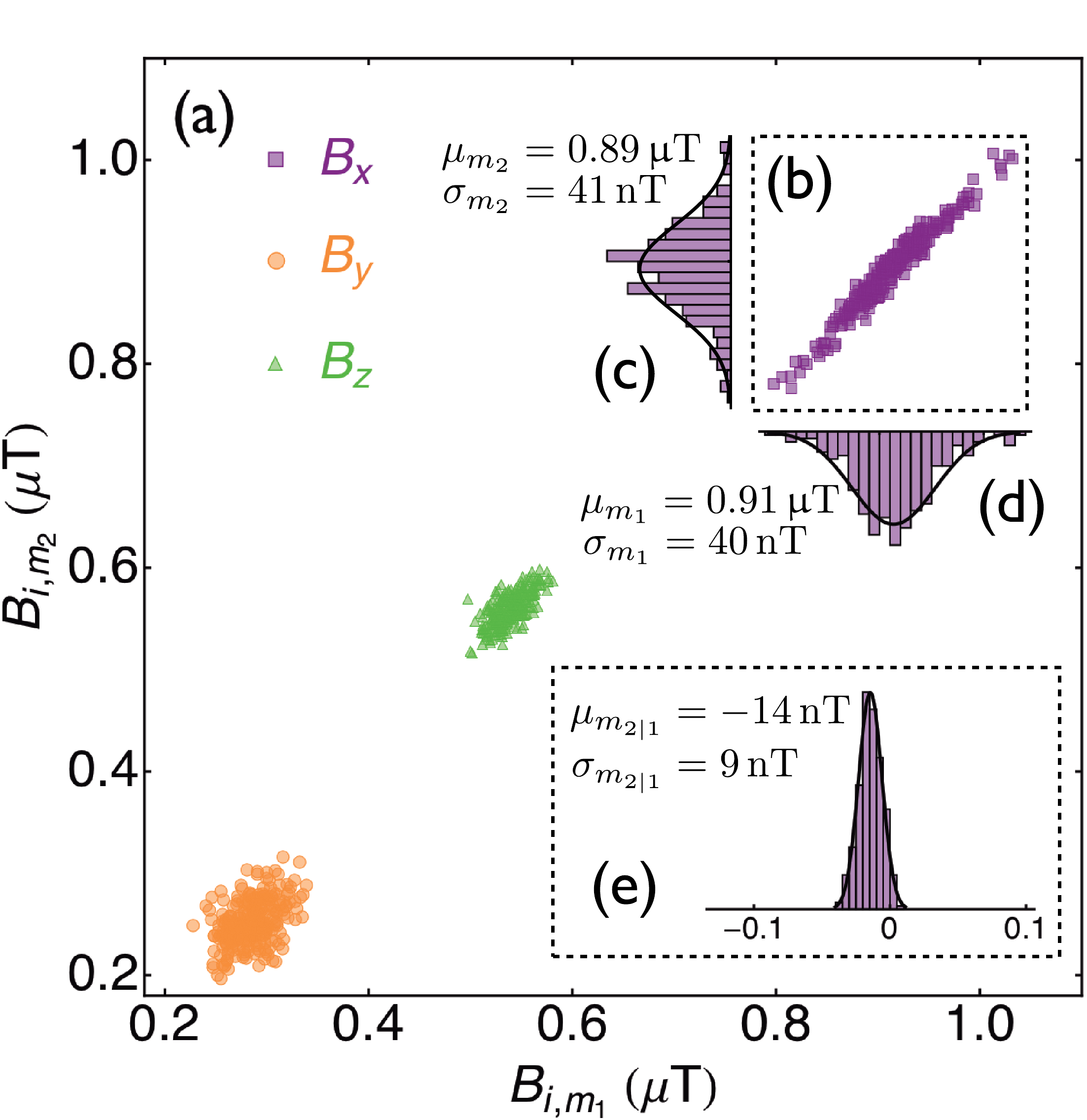}
\caption{
(Color online)
(a) Correlation plot of the field components $B_i$ estimated from consecutive measurements $m_1$ and $m_2$ of the field.
Marginal distributions from the $B_x$ component (b) are shown in histograms (c) and (d).
Each has a standard deviation of $\sigma_{B_x}\simeq\unit{40}{\nano\tesla}$.
Solid lines are gaussians with the indicated mean and standard deviation, suggesting that the field noise is approximately normally distributed.
The conditional uncertainty of the measurement $m_2$ given the outcome of measurement $m_1$ is equivalent to the dispersion of the residuals of a linear regression of $m_2$ on $m_1$, shown in inset (e), and has a standard deviation $\sigma_{m_2|m_1}=\unit{9}{\nano\tesla}$.
\label{fig:BXcorrelationPlot}}
\end{figure}

We are interested in our ability to predict (or retrodict) the magnetic field at a moment shortly after (or before) the magnetic field measurement.  This ability determines the precision of corrections for, or control of, the field seen by the atoms. To quantify this precision, we analyze the conditional uncertainty between consecutive measurements $m_1,m_2$ at $T={29}{},~ \unit{30}{\milli\second}$, shown in Figure \ref{fig:equivalentTimeSample}, inset.
Typical experimental data are shown in Figure~\ref{fig:BXcorrelationPlot}.
For a single parameter, the conditional variance is $\var(y|x)\equiv\var(y-a x)$, where the {correlation parameter} $a\equiv\cov(x,y)/\var(x)$ minimizes the conditional variance.
This is equivalent to minimizing the residuals of a linear regression $y=ax+b$, and is illustrated for a single parameter in Figure~\ref{fig:BXcorrelationPlot}(b)--(e).
For the data shown, the conditional uncertainty is $\sigma_{m_2|m_1}=\unit{9}{\nano\tesla}$.

This analysis is readily extended to multivariate data.
If $\x$ and $\y$ are vectors of parameters, with covariance matrices $\Gamma_{\x}\equiv\cov(x_i,x_j)$,  and $\Gamma_{\x,\y}\equiv\cov(x_i,y_j)$, then the conditional covariance matrix is given by
\begin{equation}
\Gamma_{\y|\x}=\Gamma_{\y}-\Gamma_{\y,\x}\Gamma_{\x}^{-1}\Gamma_{\x,\y}.
\label{eq:ConditionalCovarianceMatrix}
\end{equation}
The matrix of coefficients $A=\Gamma_{\x}^{-1}\Gamma_{\x,\y}$ minimizes the mean squared error of the linear regression $\y=A\x+B$.\cite{Kendall1979}

For the data shown in Figure~\ref{fig:BXcorrelationPlot}, the covariance matrix for the first measurement is
\begin{equation}
\Gamma_{\x} = \left(
\begin{array}{ccc}
 1.60 & 0.21 & -0.14 \\
 0.21 & 0.42 & 0.00 \\
 -0.14 & 0.00 & 0.24 \\
\end{array}
\right) \times10^{-15} ~ {\rm T}^2.
\end{equation}
The corresponding conditional covariance matrix is
\begin{equation}
\label{eq:CondCovarMat}
\Gamma_{\y|\x} = \left(
\begin{array}{ccc}
 5.6 & -8.0 & -6.0 \\
 -8.0 & 36.0 & 1.5 \\
 -6.0 & 1.5 & 12.0 \\
\end{array}
\right) \times10^{-17} ~ {\rm T}^2.
\end{equation}
This shows strong correlations among the different $B$ components, and it is interesting to diagonalize $\Gamma_{\y|\x}$ to find uncertainties $(\delta B_1, \delta B_2, \delta B_3) = \unit{(19.5,12.0, 3.0)}{\nano\tesla}$ along the directions $\bfe_1 = (-0.26,0.96,0.11),$ $\bfe_2 = (-0.41,-0.22,0.89)$ and $\bfe_3 = (0.87,0.18,0.45)$, respectively.  
We note that $\bfe_3$ is nearly the field direction, indicating {good predictability} for the magnitude of the field.
We observe similar results if we analyze the correlation between two measurements $p_i$ at the same phase on different cycles of the \unit{50}{\hertz} mains line.  It should be noted that these results include readout noise, which we now compute. 

Faraday rotation measurement at or near the shot-noise limit has been demonstrated with a variety of cold atom systems, including released MOTs, \cite{GeremiaPRA2006,StocktonThesis2007,TakanoPRL2009} optical lattices,\cite{SmithJPB2011} and optical dipole traps.\cite{HigbiePRL2005,LiuPRL2009} Our experiment is shot-noise limited by \unit{10}{dB} at 10$^7$ photons/pulse.\cite{KoschorreckPRL2010a, SewellPRL2012}  
We compute the shot-noise-limited sensitivity using Fisher Information (FI) theory.\cite{FisherMPCPS1925}
For a normally-distributed random variable $\tilde{\theta}$ with fixed variance $\sigma^2$ and mean $\theta(\bunk)$, where $\bunk$ is a vector of parameters, the FI matrix is ${\cal I}_{ij} =   \sigma^{-2} [ \partial_i {\theta} ] \partial_j {\theta}$, where $\partial_a$ represents $\partial/\partial x_a$.  
This directly gives the covariance matrix for $\bunk$ as $\Gamma_\bunk = {\cal I}^{-1}$.  
Due to shot noise, the measured rotation angles are normally distributed with $\sigma_{\theta}^2 = 1/N_{\rm p}$ and means $\theta_{1,2}$ from Eqs. (\ref{eq:FIDSignalZStart}),  (\ref{eq:FIDSignalYStart}).  
Also, the FI  is additive over independent measurements, so the FI matrix from FID is 
\begin{equation}
{\cal I}_{ij} = N_{\rm p}  \sum_{k,l} [ \partial_i \theta_l(t_k) ] \partial_j \theta_{l}(t_k).
\end{equation}
where $\bunk \equiv [B_x, B_y, B_z, \ttwo, F_z(0), F_y(0)]$ and $\{ t_k \}$ are the measurement times.  

Considering $\gamma =  \unit{2 \pi \times 7 }{\giga\hertz\per\tesla} $ for the ground states of $^{87}$Rb, and typical values from the data of Figure~\ref{fig:BXcorrelationPlot}: $(B_x,B_y,B_z)= \unit{(910,285,540)}{\nano\tesla}$, $\ttwo = \unit{1.48}{\milli\second}$, $F_z(0) = F_y(0) = \unit{10^5}{spins}$, $G = \unit{0.89\times10^{-7}}{\radian\per spin}$ and $N_{\rm p} = \unit{10^7}{photons}$, we find the covariance matrix ({\bf B} portion only) 
\begin{equation}
\Gamma_{\rm SN} = \left(
\begin{array}{ccc}
 1.30 & -2.43 & -1. 00 \\
 -2.43 & 11.87 & -1.61 \\
 -1.00 & -1.61 & 2.57 \\
\end{array}
\right) \times10^{-17} ~ {\rm T}^2.
\end{equation}
If we diagonalize $\Gamma_{\rm SN}$ we find uncertainties $(\delta B_1, \delta B_2, \delta B_3) = \unit{(11.2, 5.6, 0.6)}{\nano\tesla}$, along the directions $\bfe_1 = (0.2,-0.97,0.14), \bfe_2 = (0.5,-0.02,-0.86)$ and $\bfe_3 = (0.84,0.24,0.48)$, respectively.

We can now correct the measured field noise of Eq. (\ref{eq:CondCovarMat}) for the measurement noise, to find the field noise $\Gamma_{\rm B} = \Gamma_{\y|\x} - \Gamma_{\rm SN}$ of
\begin{equation}
\Gamma_{\rm B} = \left(
\begin{array}{ccc}
 4.30 & -5.50 & -5.00 \\
 -5.50 & 24.00 & 3.10 \\
 -5.00 & 3.10 & 9.50 \\
\end{array}
\right) \times10^{-17} ~ {\rm T}^2
\end{equation}
or $\delta B_i\simeq\unit{10}{\nano\tesla}$ integrated over the \unit{\kilo\hertz} bandwidth of the measurement.

The FI analysis also reveals that $\delta F_z(0)$ and $\delta F_y(0)$, the noises in the atomic state preparation, are only very weakly coupled into the estimates of ${\bf B}$ and $\ttwo$, making the measurement insensitive to, e.g., atom number fluctuations and variation in the optical pumping efficiency. 

We have demonstrated a cold-atom magnetometer with sub-nT sensitivity, \unit{20}{\micro\meter} transverse spatial resolution and \unit{1}{\milli\second} temporal resolution.  The instrument gives simultaneous information about the three field components plus one gradient component and requires no additional applied fields, making it very attractive for non-disturbing field monitoring and control.  We note that sensitivity can be improved by increasing the number of atoms (in our system a five-fold improvement to $10^6$ atoms is readily achievable\cite{KubasikPRA2009}) and/or the number of photons, although tensor light shifts should be taken into account for larger photon numbers.\cite{SmithPRL2004,StocktonThesis2007,KubasikPRA2009}

\begin{acknowledgments}
This work was supported by the Spanish MINECO under the project MAGO (Ref. FIS2011-23520), by the European Research Council  project {AQUMET} and by Fundaci\'{o} Privada Cellex.
\end{acknowledgments}

%

\end{document}